\documentclass[pre,floatfix,notitlepage,nofootinbib,superscriptaddress]{revtex4-1}

\usepackage{bm,amsmath,amssymb,graphicx}
\usepackage{ amsfonts, bm, epsfig, xcolor}   

\usepackage{textcomp} 

\usepackage[format=plain,labelfont={bf,small},textfont=small,justification=raggedright,singlelinecheck=false]{caption}

\begin{document}

\title{Buckling mediated by mobile localized elastic excitations}
\author{R. S. Hutton}
	\email{rhutton@unr.edu}
\affiliation{Department of Mechanical Engineering, University of Nevada, 1664  N.  Virginia  St.\  (0312),  Reno,  NV  89557-0312,  U.S.A.} 
\author{E. Vitral}
	\email{vitralfr@rose-hulman.edu}
\affiliation{Department of Mechanical Engineering, Rose-Hulman Institute of Technology,
5500 Wabash Ave., Terre Haute, IN 47803, U.S.A.} 
\author{E. Hamm}
	\email{luis.hamm@usach.cl}
\affiliation{Departamento de F\'isica, Facultad de Ciencia, Universidad de Santiago de Chile, Av.\ V\'ictor Jara 3493, Estaci\'{o}n Central, Santiago 9160000, Chile}
\author{J. A. Hanna}
	\email{jhanna@unr.edu}
\affiliation{Department of Mechanical Engineering, University of Nevada, 1664  N.  Virginia  St.\  (0312),  Reno,  NV  89557-0312,  U.S.A.}

   \begin{abstract}
Experiments reveal that structural 
transitions in thin sheets are mediated by the passage of transient and stable mobile localized elastic excitations. 
These ``crumples'' or ``d-cones'' 
nucleate, propagate, interact, annihilate, and escape. 
Much of the dynamics occurs on millisecond time scales. 
Nucleation sites correspond to regions where generators of the ideal unstretched surface converge. 
Additional stable intermediate states illustrate two forms of quasistatic inter-crumple interaction through ridges or valleys.
These interactions 
create pairs from which extended patterns may be constructed in larger specimens.
The onset of localized transient deformation with increasing sheet size 
 is correlated with a characteristic stable crumple size, whose measured scaling with thickness is consistent with prior theory and experiment for localized elastic features in thin sheets.  We offer a new theoretical justification of this scaling. 
\end{abstract}

\date{\today}

\maketitle

\section{Introduction}
Localization of strains in continua can bypass a more energetically expensive homogeneous deformation route, mediating 
 global changes by the passage of small increments of shear, slip, or some geometric feature such as curvature orientation. 
Established and speculative examples of mechanisms involving mobile localized regions include 
plastic deformation by microscopic objects such as dislocations 
 in crystals \cite{Hirth85}, shear transformation zones in glasses \cite{FalkLanger98}, or diffusing free volume in granular packings \cite{Bazant06}, 
reversible and irreversible topological restructuring events in foams \cite{Durian97}, 
and continuum-level phenomena such as elastic surface waves at sliding contacts on soft materials \cite{Schallamach71, Viswanathan16}, or 
 L{\"{u}}ders bands in metallic alloys 
and necks in cold drawn polymers \cite{Nadai50}. 
Examples most related to the present work are the appearance of localized features during the intermediate stage of wavelength-refinement of a buckled sheet compressed within an arch-shaped gap \cite{RomanPocheau12}, and propagating one-dimensional elastic and plastic instabilities triggered by the birth of indentations or bulges in pressurized elongated shells, which serve to transform the structure from a metastable to a stable state \cite{Kyriakides94}. 

Elastic stretching and bending energies are known to localize in thin structures such as sheets and shells under quasistatic loading and confinement, forming crescent-shaped features of an extended dipolar-Gau{\ss}ian character \cite{BenAmarPomeau97, Witten07},  
 which we will here refer to as ``crumples''.\footnote{This terminology somewhat follows \cite{King12, Timounay20}, although the latter use the term to refer to what we would call a valley-bound pair of crumples. A more common, but somewhat controversial, term is ``d-cone'' \cite{BenAmarPomeau97}.} 
Researchers have examined the conditions for their nucleation under direct indentation \cite{PauchardRica98, MoraBoudaoud02, Das07, Potter07, Nasto13} or more complex loading and frustration \cite{DiDonna02, Horak06, Schroll11, King12}, 
their sequential induction \cite{Hunt20, Marthelot17}, 
and their status as localized units, related through snaking bifurcations to extended oscillatory patterns \cite{Thompson15, Hunt00, Hunt20}  
of a type observed across a range of scales \cite{Harris61, Timounay20}. 
To our knowledge, the only prior work on interaction of mobile crumples has been in the quasistatic regime \cite{Hamm04, Walsh11}. 
While reminiscent of more familiar singular solutions of linear field theories, such as point charges or vortices, crumples cannot simply be superposed, as the governing equations are nonlinear.
Nonetheless, we might seek some type of reduced description of buckled structures centered on these elastic excitations, taking into account their nucleation, motion, and interaction with each other and a background curvature field. 

In this initial report, 
we reveal that shape changes in 
 extended thin elastic objects are mediated by the nucleation and transit of crumples, even in dynamic settings when these localized structures are unstable.
We employ a simple experimental configuration to induce multi-stable buckling and snap-through in bent sheets of varying aspect ratio, and document several types of rapid transient and quasistatic events.
Crumples are seen to exhibit two fundamental types of stable pair interaction, related to the organization of larger-scale patterns \cite{crumpledynamics}.


\begin{figure*}[h!]
    \includegraphics[width=\linewidth]{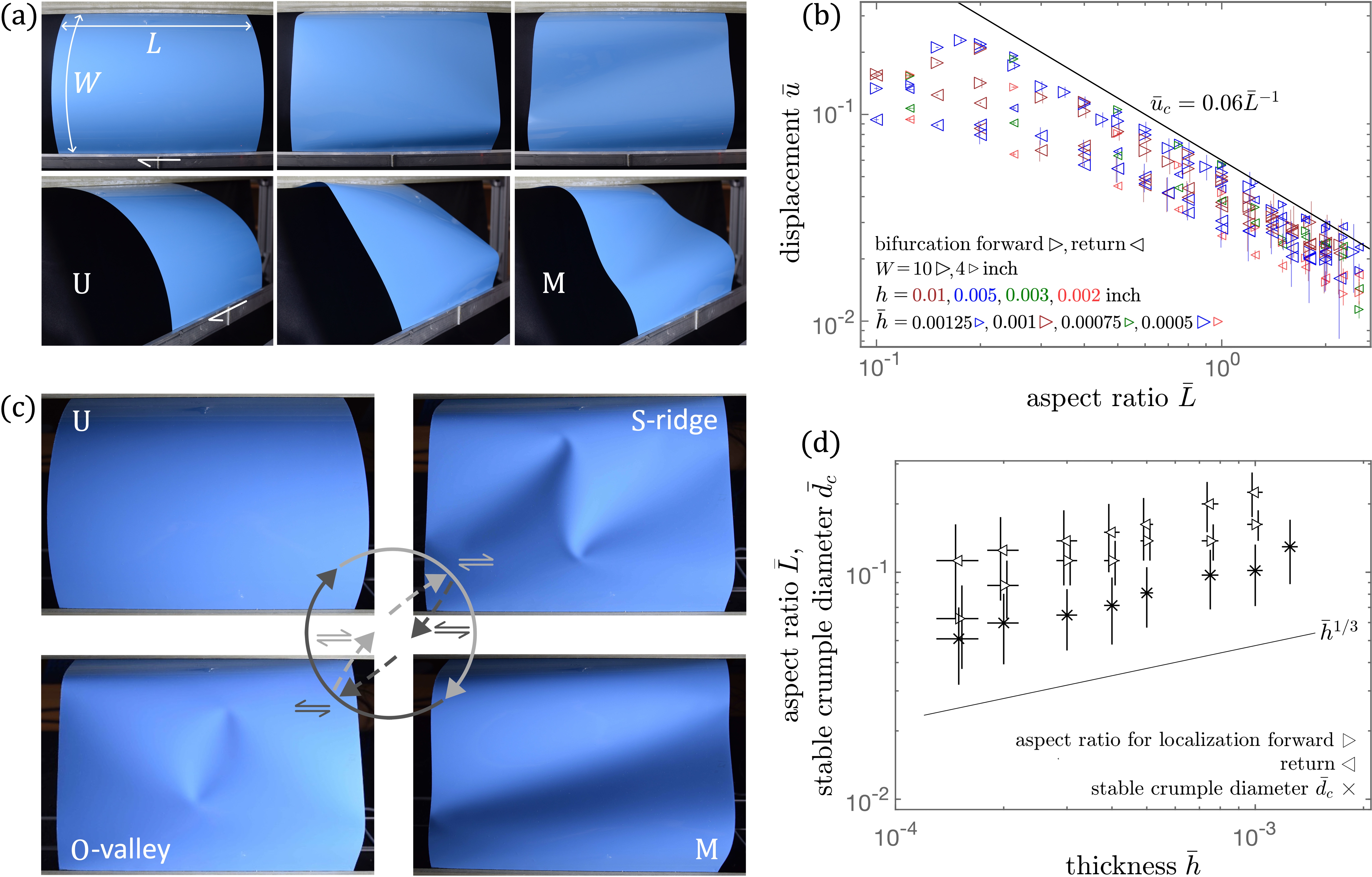}
   \caption{
   (a) Sheets of width $W$ (an overbar will denote normalization by this quantity) and length $L$ are clamped in a reference quarter-cylinder U state.  A parallel lateral displacement $u$ is applied to a flat side, leading to formation of a ridge and sharpening of features, then snap-through to the M state. 
    (b) Critical displacement $\bar{u}$ for bifurcation away from the U state (forward) and M state (return) as a function of the aspect ratio $\bar{L}$, for several sheet thicknesses $h$ and widths $W$. Some error bars are horizontally shifted for clarity.  The theoretical line $\bar{u}_c= 0.06\bar{L}^{-1}$ is when the sharpening of features in the U state terminates in an ``origamized'' limit (Appendix \ref{origami}); this has no fitting parameters, but uses assumptions valid for large $\bar{L}$. 
   (c) Principal hysteresis loop in an $\bar L = 1.2$, $\bar h = 5\times 10^{-4}$ sheet: forward shear from U to M \emph{via} stable S-ridge, return shear from M to U \emph{via} stable O-valley (see also \cite[V2]{vidsite}).   Paths between the two intermediate states involve other states (not shown, see also \cite[V3]{vidsite}). 
   (d) Crumple diameter $\bar{d}_c$ (measured as described in Appendix \ref{crumplemeasure}), and minimum aspect ratio $\bar{L}$ for localization during the forward and return bifurcations, as a function of the thickness $\bar{h}$. 
   Some error bars are horizontally shifted for clarity. 
   }
    \label{composite}
\end{figure*}

\section{Overview of experiments}

Thin elastic sheets of thickness $h$ are bent into a quarter-cylinder configuration with a free-span width $W$ (hereafter, an overbar will denote normalization of a quantity by $W$) and length $L$, their flat sides attached to rigid parallel beams. Details are provided in Appendix \ref{errors}. 
The sheets are quasistatically ``sheared'' by parallel lateral displacement $u$ of one beam, to effect an elastic snap-through transition between unimodal ``U'' 
 (quarter-cylinder, $u=0$) and multimodal ``M'' configurations. 
The generic features of this transition are easiest to see in larger aspect ratio plates (Figure \ref{composite}(a)).  As the U state is sheared forward, mean curvature concentrates in peaks near the compressive (trailing) corners, with 
something reminiscent of a Witten ridge \cite{WittenLi93, Lobkovsky95} between them. For narrow plates, this ridge is comparable to the plate size, giving the appearance of a synclastically curved beam.  
The M state features a long valley between the tensile (leading) corners. We refer to ridges and valleys with respect to the initial sense of mean curvature of the quarter-cylinder. 
These initial and terminal states are potentially developable, which would entail no stretching of the sheet.  The present study pertains to observations that transitions between these two states, and any intermediate stable states, are mediated by the passage of transient and/or stable crumples--- localized excitations of concentrated stretching and bending energy. 
Crumples generally present as indentations oriented with their positive intrinsic curvature pointing towards a clamped edge near one of the tensile corners.
When in isolation, crumples glide so as to follow the imposed shear, but this is often not the case when interacting with other crumples. 
A crumple transports mean curvature, so that its passage through the sheet effects a global change between a ridge and a valley.

\section{Results}
\subsection{Snap-through}

The initial bifurcation away from the U branch in forward shear is well-predicted for large aspect ratios $\bar{L}$ by a purely geometric, thickness-independent model with no fitting parameters. The body is assumed to evolve along a U-like branch until it ``origamizes'' in a sharp-cornered configuration beyond which it must either stretch, or jump to another branch of solutions, which may correspond either to partial or complete snap-through.
Details of the trigonometric calculation and its underlying assumptions are provided in Appendix \ref{origami}; the result is plotted in Figure \ref{composite}(b) alongside experimental data for the normalized displacement $\bar u$ at forward and return bifurcations as a function of aspect ratio $\bar L$ for
several thicknesses $\bar h$. 
Data at the very lowest aspect ratios correspond to homogeneous, rather than localized, deformations.
The transition from M to U upon return shear is triggered by loss of stability of the M state, which is not expected to correspond to any particular origamized limit, and may be related to the corresponding transition for narrow bands \cite{YuHanna19}. 
Yet there is also no obvious indication of thickness dependence in this data.
For intermediate aspect ratios, there is a bistable range of displacement, in which a ridge oriented towards compressive corners in the U state may be exchanged for a valley oriented towards tensile corners in the M state. At the largest aspect ratios, there is a range of displacement in which neither U nor M is stable.

\subsection{Transients and nucleation} 

We present a general overview of transient phenomena using results for sheets of $\bar h = 5\times 10^{-4}$. 
Using a fast camera, we confirm that snap-through in narrow strips ($\bar{L} \le 0.15\pm0.025$) is homogeneous, with no noticeable localization \cite[V1 0:00]{vidsite}. 
At larger aspect ratios, we observe localization in the form of one or more transient crumples.  These events are too fast for the naked eye to detect.

\begin{figure*}[h]
    \includegraphics[width=\linewidth]{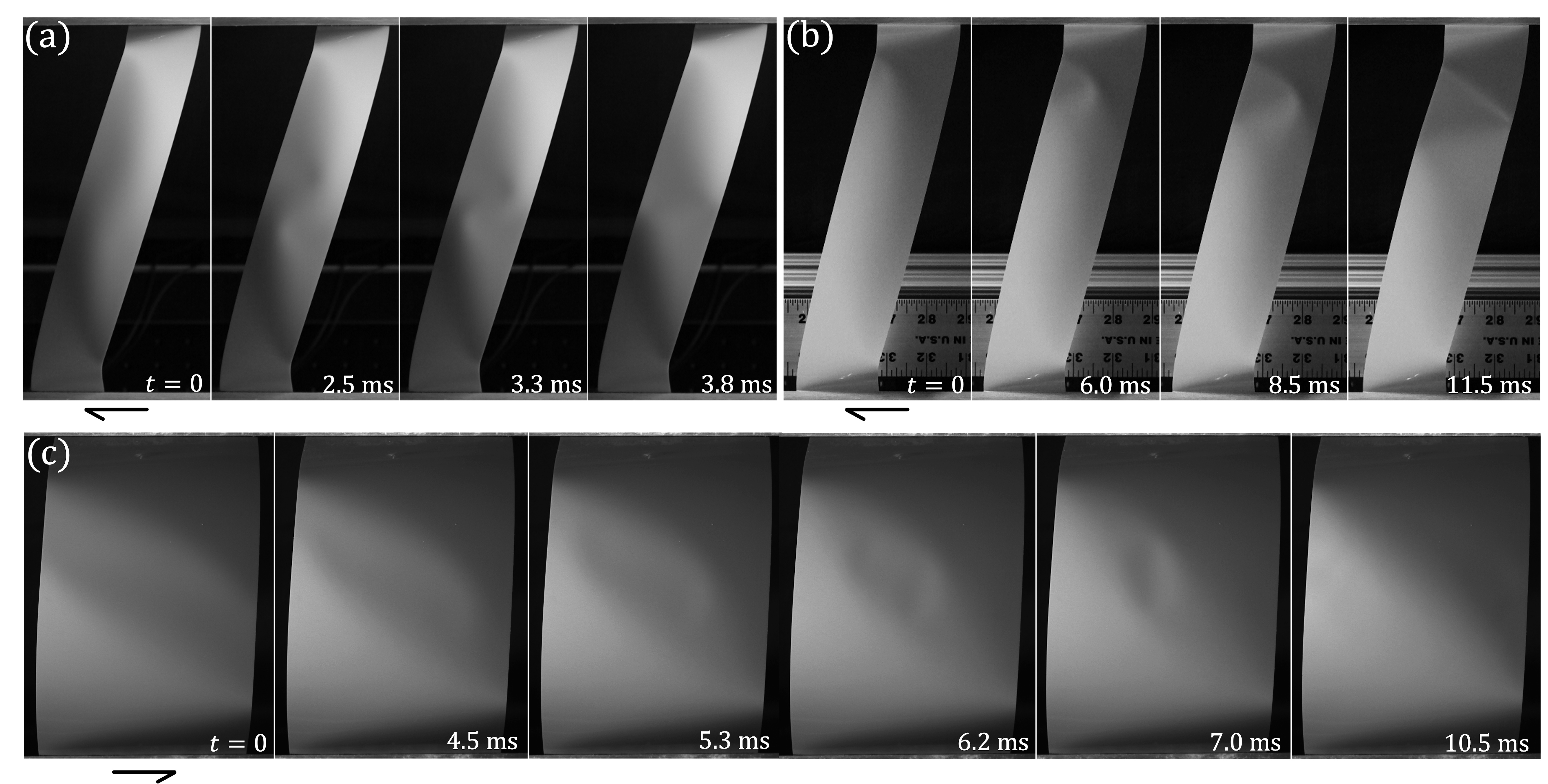}
   \caption{Transient phenomena mediating localized snap-through of narrow sheets of thickness $\bar h = 5\times 10^{-4}$. 
   (a) Aspect ratio $\bar L = 0.175$, U to M transition: a pair of crumples nucleate near the central ridge of the strip and take a nonlinear path out through the sides (6000 fps, see also \cite[V1 0:08]{vidsite}).
   (b) $\bar L = 0.2$, U to M transition: a single crumple nucleates in a highly curved corner, the edge of a ridge, and travels to the other side  (2000 fps, see also \cite[V1 0:19]{vidsite}). 
   (c) $\bar L = 0.6$, M to U transition: shallow crumples nucleate near inflection points on the sides, travel inward and imperfectly annihilate (4000 fps, see also \cite[V1 0:42]{vidsite}). 
}
    \label{transientsnarrow} 
\end{figure*}

In forward shear (from U to M) at $\bar{L} = 0.175$, two crumples nucleate simultaneously near the center of the strip. These follow a nonlinear path and escape from opposite free sides 
(Figure \ref{transientsnarrow}(a) and \cite[V1 0:08]{vidsite}). 
This particular behavior is only observed for a narrow range of aspect ratios. 
At $\bar{L} = 0.2$, single crumples nucleate in a highly-curved region near a compressive corner and propagate through the sheet to exit on the other side 
(Figure \ref{transientsnarrow}(b) and \cite[V1 0:19]{vidsite}). 
These corner nucleation sites are located near the conical singularities of an idealized developable surface whose generators converge just outside the sheet.
For larger aspect ratios (around $\bar{L} \ge 0.5$), there is often sufficient time for both compressive corners generate a crumple.  These may interact strongly enough to exit together through the same side of the sheet, against the tendency, in isolated motion, of one of the crumples. 
A crumple may split and partially exit from both free edges \cite[V1 0:29 and 0:50]{vidsite}. When unimpeded, these crumples move at speeds of 150-250 inch/sec, consistent with expectations for dispersive bending waves in these elastic sheets. 
At aspect ratios starting around unity, a more complex transient is observed in which crumple pairs nucleate sequentially in a high curvature region over a period of a few milliseconds  \cite[V1 1:00]{vidsite}; 
 in sufficiently large sheets this sequence may lead to stable states \cite[V1 1:24]{vidsite}, as described below.  
 The initial pair nucleation is reminiscent of the buckling of a Witten ridge \cite{DiDonna02}, while the subsequent nucleations are reminiscent of propagating instabilities \cite{Kyriakides94} and of sequential induction of stable valleys in inflated bags \cite{Timounay20} and pressurized shells \cite{Marthelot17}.

In return shear (from M to U) at $\bar{L} = 0.2$ and larger, two shallow crumples nucleate almost simultaneously near the inflection points of the free boundary, travel inwards and collide near the center.  These smaller crumples move at speeds of 650-750 inch/sec, about 3-4 times faster than the more prominent crumples observed during forward shear.
  The collision is not entirely head on, but the interaction is strong, such that the crumples usually nearly annihilate rather than pass through each other, shedding a bit of energy out through the free sides
(Figure \ref{transientsnarrow}(c) and \cite[V1 0:42]{vidsite}).  
While these nucleation sites are rather flat, they also correspond to singularities of an idealized developable surface whose generators converge just outside the sheet \cite{Korte11}.  In contrast to the forward process, the nucleation of crumples upon return occurs during relaxation, not sharpening, of these features. 
At larger aspect ratios, stable puckers appear near the tensile corners. These are sensitive to boundary conditions set by alignment, but are practically unavoidable for large sheets; they might be a result of compressive stresses transverse to the tensile diagonal, or some other local instability. 
The rapid passage of the edge-nucleated shallow crumples will knock these puckers towards the center of the sheet or out through a free side; renucleation from a free side has also been observed.  The disturbed puckers may annihilate \cite[V1 1:14]{vidsite}, or form stable states \cite[V1 1:43]{vidsite}, as described below.

\subsection{Thickness dependence of the aspect ratio for localization} 

The smallest aspect ratios $\bar{L}$ at which transient localization is observed in both forward and return shear are presented in Figure \ref{composite}(d) for about an order of magnitude range of thickness, $1.5 \times 10^{-4} \le \bar{h} \le 1 \times 10^{-3}$. 
This onset of localization is seen  (Figure \ref{composite}(d))
 to be correlated with the diameter $\bar{d}_c$ of crumples measured over the slightly larger range, $1.5 \times 10^{-4} \le \bar{h} \le 1.25 \times 10^{-3}$, which is consistent with a thickness scaling of $\bar{h}^{1/3}$. Details of the crumple size measurement are provided in Appendix \ref{crumplemeasure}.
Such a scaling for crescent singularities, first shown in experiments \cite{Cerda99} and later supported by theory and numerics \cite{CerdaMahadevan05, LiangWitten05, Liang08}, echoes that of other stretching features such as the Witten ridge \cite{WittenLi93, Lobkovsky95}. 
This correlation indicates that the sheet needs to have sufficient room--- about 1.5 to 2 crumple diameters--- for a transient crumple pair to nucleate. The sheet width is sufficiently large that finite size effects are expected only from the lateral dimension. 
The relevance of static scaling to transient behavior suggests that even transient propagating crumples may have a characteristic size.  

We offer a new argument for a thickness scaling of $1/3$ that should apply when crumples nucleate from a stretching ridge, such as our forward shear experiments or DiDonna's numerical end-compression of a Witten ridge \cite{DiDonna02}.  
Let us characterize the crescent-shaped dipolar crumple by a diameter $d_c$ and a width $w_c \ll d_c$, the latter length representing the extent of high curvature across the crumple.  The dominant contributions to the mean and intrinsic curvatures are, respectively, $H \sim w_c^{-1}$ and $K \sim d_c^{-1}w_c^{-1}$, both distributed over an area $A_c \sim d_c w_c$. 
The characteristic stress $\sim w_c^2 K$, as gleaned from the normal force balance in a plate \cite{LandauLifshitz86}. 
The dominant contributions to the total stretching and bending energies are, respectively, $E_s \sim hA_c (w_c^2K)^2$ and $E_b \sim h^3 A_c H^2$; we neglect, among other things, the smaller contribution $h^3A_cK$ to the bending energy. 
Minimizing the sum of stretching and bending energies with respect to $w_c$ yields the scaling $w_c \sim (hd_c)^{1/2}$ and thus $E_s \sim E_b \sim h^{5/2}d_c^{1/2}$.  Balancing these against the total energy of a Witten ridge, from the analysis of Lobkovsky and co-workers \cite{Lobkovsky95}, $E_r \sim h^{8/3}$ leads to the prediction $d_c \sim h^{1/3}$.  
This argument does not account for dependence on other parameters, such as the background mean curvature, which differ between our forward and return shear experiments, and does not seem to directly apply to the latter, although the same thickness scaling is observed.

\subsection{Stable crumple pairs: S-ridges, O-valleys, and their antecedent transients}

At about unit aspect ratio or larger, the structure displays additional quasistatic transitions through stable intermediate states.
We refer here to certain stable states reached by boundary shearing of a sheet around a principal hysteresis loop from U to M and back (Figure \ref{composite}(c) and \cite[V2]{vidsite}). 
The range of stability of these and other states, as generated by direct external forcing of the sheet, is greater than that accessible by boundary shearing. Other states, some of which appear in \cite[V3]{vidsite}, may be reached through an internal shearing loop between intermediate states.
A stable O-valley can be induced in the U state, and a stable S-ridge in the M state, by external poking, at aspect ratios smaller than those at which these pairs arise dynamically through boundary shearing.
While the structures can sustain single crumples, 
the two states of greatest interest involve two crumples paired in one of two distinct ways. 
In one state, which we call an S-ridge, the two crumples are connected through a shared ridge. 
When a single S-ridge sits in the center of the sheet, the crumples' valleys extend out to opposite free sides.
In the other state, which we call an O-valley, the two crumples are connected through a shared valley,  oriented towards the tensile direction and pointing towards the clamped edges of the sheet\footnote{This valley-bound pair of crumples is what \cite{Timounay20} call a ``crumple''. Furthermore, what \cite{Hamm04} call an ``S-shaped ridge'' in a twisted cylindrically-bent sheet is unrelated to our S-ridge, and instead resembles a well-separated, curved O-valley pair.  These are also visible in the early torsion experiments of \cite{Harris61}.} (Figure \ref{composite}(c)).  
Note that the difference between a ridge bond and a valley bond is not simply due to the bias of the bent sheet--- the ridge and valley are two distinct portions of the crescent singularity.  
Other, weaker ridges connect these pairs to the high-mean-curvature regions near the compressive edges of the sheet.  These weaker bonds are relevant to longer chains of crumple pairs, a topic reserved for a subsequent publication \cite{crumpledynamics}. 
Like transient crumples, static crumples also require sufficient space in the sheet to form, a characteristic that depends on the sheet thickness.
The energy landscape is such that crumples feel short-range attraction to other crumples and boundaries, and so must be sufficiently far from either in order to be in a metastable equilibrium such as a valley pair.  Further exploration of this landscape will be addressed in a subsequent publication.

\begin{figure*}[h]
    \includegraphics[width=\linewidth]{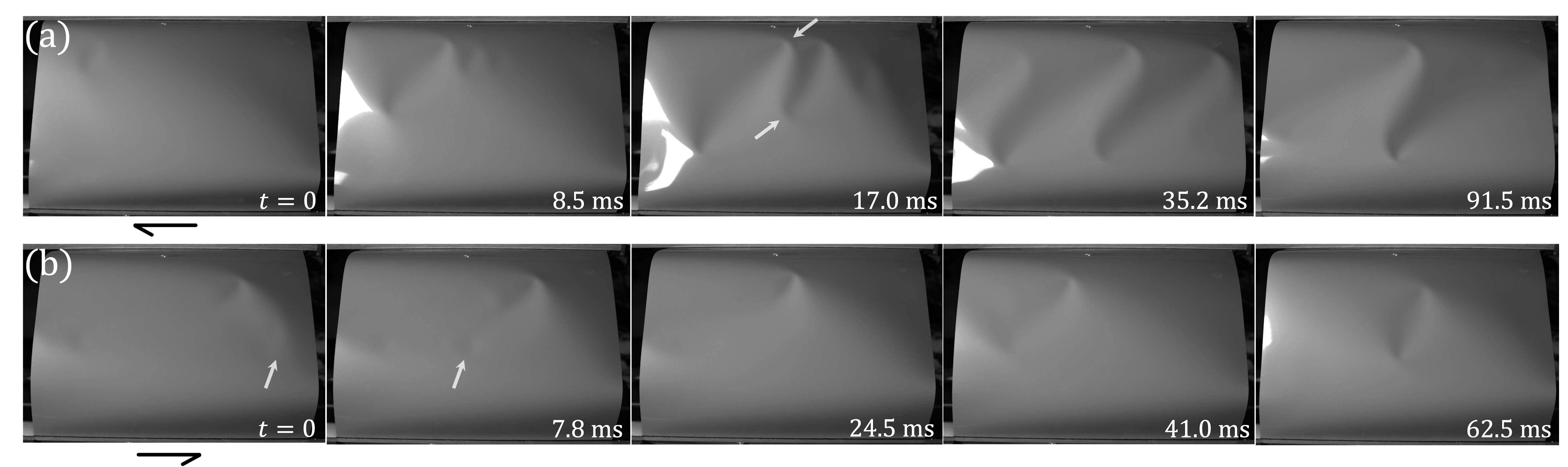}
  \caption{
  Complex transient phenomena leading to stable crumple pairs in larger sheets of aspect ratio $\bar L = 1.2$, thickness $\bar h = 5\times 10^{-4}$. Crumple positions continue to oscillate after the last frames shown here. 
  (a) In the U to M transition, events begin with nucleation of a transient valley near a corner, which triggers a cascade. Two (arrowed) crumples eventually remain to form an S-ridge (4000 fps, see also \cite[V1 1:24]{vidsite}). 
  (b) In the M to U transition, stable puckers are initially present. A shallow crumple (arrowed) nucleated near an inflection point begins a series of events leading to a stable O-valley (4000 fps, see also \cite[V1 1:43]{vidsite}). 
 }
  \label{transientswide}
\end{figure*}

In the principal loop we consider, forward shear from U to M is mediated by a stable S-ridge that tilts and extends a bit under increasing shear, rotating slightly towards the tensile direction, then exits as a pair through one of the free sides of the sheet to complete the process. 
Return shear from M to U is mediated by a stable O-valley that tilts and shrinks a bit under decreasing shear, rotating slightly away from the tensile direction, and then collapses as the two crumples annihilate each other to complete the process \cite[V2]{vidsite}. 
The transient phenomena giving rise to these intermediate states are complex, occurring over a time of less than 100 ms, not including a comparable or longer period of oscillation that follows. 
In forward shear, the propagating nucleation of transient valleys results in the eventual pairing of a crumple from each of the two largest initial valleys to form a stable S-ridge, with other crumples being expelled from the sides of the sheet (Figure \ref{transientswide}(a) and \cite[V1 1:24]{vidsite}). 
In return shear, the potential for stable puckers to slowly grow into crumples is interrupted by fast transient passage of shallow crumples nucleated at the free sides, which knock the puckers towards the center of the sheet.  
The end result is formation of a stable O-valley from the disturbed pucker-crumples, sometimes with renucleation of a pucker-crumple at the edge, after the passage of the fast shallow crumples (Figure \ref{transientswide}(b) and \cite[V1 1:43]{vidsite}). 
None of the transient behaviors in forward or return shear are observable in real time.


\subsection{Additional observations}

A few other observations are collected and demonstrated in \cite[V3]{vidsite}. 
Under sufficiently slow steady loading, we can observe noticeable jumps as two crumples bind into, or unbind out of, an O-valley configuration.
The S-ridge configuration is strongly bound (attempted unbinding can even lead to nucleation of a second S-ridge in between \cite{crumpledynamics}), but highly mobile as a pair. 
Single isolated crumples appear to induce a small pucker of opposite curvature nearby. 
New crumples are sometimes induced at free edges by the approach of a crumple.  This can happen, for example, after separating the crumples in an O-valley such that two S-ridges are obtained.

\section{Conclusions and future directions}

In conclusion, we have reported several hitherto unremarked dynamic phenomena and pair interactions in the structural transitions of a bent sheet, revealing that such transitions are inhomogeneous, mediated by the passage of unstable and stable localized structures. 
These nucleate reproducibly at regions of converging generators, requiring sufficient space that scales with thickness in a manner consistent with new and prior theory and observations of a characteristic crumple size.
It is likely that the complex behaviors observed by probing this simple system are representative of plate and shell structural dynamics under generic loading.   
Our next report will examine larger-scale patterns formed from the two fundamental crumple pairs described here.  It will be of interest to infer the associated nucleation and binding energies from force and curvature measurements.

\section*{Acknowledgments} 
We thank N. Corbin for serendipitous camera pointing, S. Riddle and T. Yu for early experimental exploration, D. Gimeno and J. Rudfelt for experimental assistance, and R. Groh, S. Kyriakides, and C. D. Santangelo for discussions. 
JAH and EH thank the organizers of the Variational Models of Soft Matter conference in Santiago de Chile for their indirect role in initiating this collaborative work. 
We acknowledge support from the National Science Foundation under grant CMMI-2210797 and from the National Aeronautics and Space Administration through the Nevada Space Grant Consortium under grant 80NSSC20M00043.

\appendix

\section{Snap-through measurements and errors}\label{errors}

Thin elastic sheets (polyester shim stock, Artus Corp., Englewood, NJ)
 of thicknesses ${0.0015 \pm 0.0002} \le h \le {0.0125 \pm 0.0005}$ inch are bent into a quarter-cylinder configuration with
  either a fixed free-span width of $W = 10$ inch and varying lengths $1 \le L \le 24$ inch or a 
free-span width $W = 4$ inch and lengths $0.5 \le L \le 10$ inch, their flat sides attached to rigid parallel beams 
 using double-sided rubber adhesive tape 
and additional mechanical clamps near the corners. As the material is mildly anisotropic, all data were obtained using sheets with the machine direction 
oriented parallel to the beams. 
The testable range is limited by gravitational slumping at low $\bar h$ and stiffness-induced clamping difficulties at high $\bar h$.

Two groups of experiments were performed, under different conditions.  

One group used $W = 10$, and comprises one of the two sets of $W = 10$ inch, $h=0.005, 0.01$ inch data for $\bar u$ at bifurcations in Figure \ref{composite}(b), along with all of the data for crumple sizes (detailed in a subsequent subsection) and $\bar{L}$ at first localization in Figure \ref{composite}(d), and all of the video footage.
 They employed a generic translation stage; quasistatic ``shear'' displacement rates were between 0.005-0.160 inch/sec, with 0.024 being typical.  
Data for $\bar u$ were obtained using an LVIT sensor (Omega Engineering, Norwalk, CT) attached to the stage. 
Uncertainty is under $\pm 0.06$ inch, most of this due to noise in the measurement circuit. 
These are single observations obtained from sequential recutting of the same samples. 
Uncertainty in $L$ from manual cutting of the sheets is 
 $\pm$ 0.016 inch.
Data for smallest aspect ratios $\bar{L}$ at first localization are for several independent observations, and the vertical error bars reflect ambiguity in the transition from homogeneous to fully localized deformation. 
The two thinnest sheets were examined in an inverted configuration to partially mitigate noticeable effects of gravity. 

A second, more accurate, group of experiments 
comprises the remaining $\bar u$ at bifurcation data, 
including sheets with $W = 4$ inch, $h=0.005, 0.003, 0.002$ inch, and the second of the two sets of $W = 10$ inch, $h=0.005, 0.01$ inch data. 
These employed output from a translation stage unit (Thorlabs, Inc., Newton, NJ)
run at a very low displacement rate of 0.05mm/sec. 
Data sets were obtained from sequential recutting of the same samples. 
The uncertainty is a combination of the stage positional accuracy $\pm$ 5 \textmu m and backlash $\pm$ 2 \textmu m, and a conservative estimate of the uncertainty, due to human error, of the symmetric U reference state alignment 
of $\pm$ 100 \textmu m for $W = 4$ inch sheets and $\pm$ 200 \textmu m for $W = 10$ inch sheets. 
  Additionally, the stage was further stabilized with supports against twist, and data were taken after ``running in'' the system by executing one trip through the principal hysteresis loop. 
 Bifurcations were observed in three subsequent loops and averaged to produce a data point, with vertical error bars reflecting the uncertainty added to the max/min observations.

Films were made at 2000-6000 frames/sec using an AX-100 camera with PFV4 software including an HDR tool (Photron USA, San Diego, CA), and an AF NIKKOR 24-85mm f/2.4-4 lens (Nikon USA, Melville, NY).

\section{Origamization prediction for bifurcation in forward shear}\label{origami}

\begin{figure}[h]
        \includegraphics[width=0.6\linewidth]{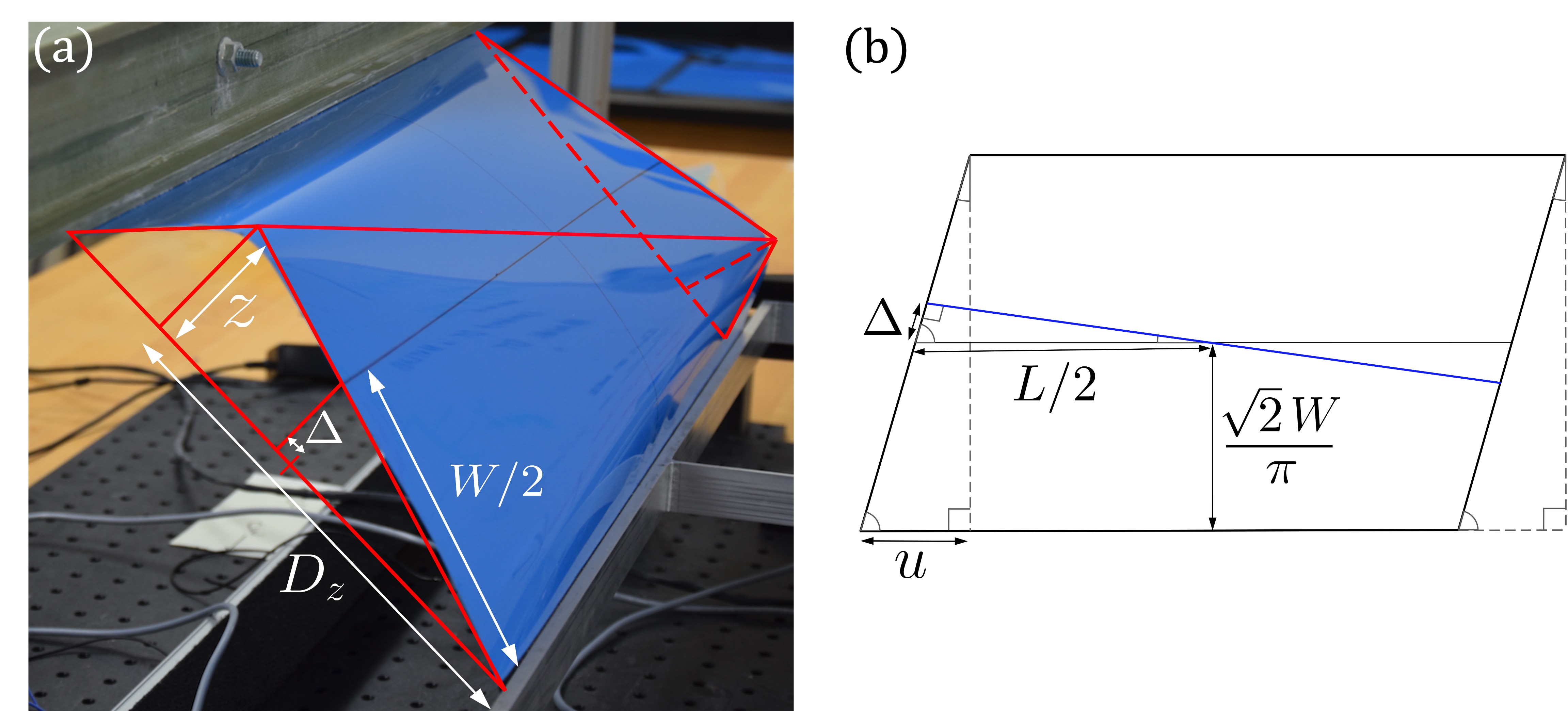}
   \caption{Geometry (a) of the sheet and (b) of its projection on to the secant plane, when the U state origamizes.   Details of the calculation are in the text.  }
    \label{trig} 
\end{figure}

The U branch is assumed to evolve isometrically until a certain limiting state is reached.
 We approximate this state using assumptions appropriate to large aspect ratios $\bar{L}$, for which the critical displacement $\bar{u}_c$ at bifurcation is small.
Consider the secant plane between the two clamped edges of the sheet, which are separated by a perpendicular distance $2\sqrt{2}\,W/\pi$.  
In the initial U state, the maximum height 
\begin{equation}
z = (2-\sqrt{2}\,)W/\pi 
\end{equation}
of the sheet above this plane occurs along a generator parallel to and midway between these edges. 
Inspired by experimental observations, we assume that this maximum height remains the same as the structure evolves under shear, until it reaches the sharp-cornered configuration of Figure \ref{trig}(a) at some critical $\bar{u}_c$.\footnote{This state is entirely distinct from the limiting state described in \cite[Appendix A]{YuHanna19}, which represents the global maximum isometric displacement the sheet may adopt, in which the generator sits flat between the tensile corners.  Note also that while the setup shares some similarities, the symbols $L$ and $W$ are switched in the present paper.}
For small $\bar{u}_c$, we may approximate the distance $D_z$ in the plane between the clamped edge and the location of maximum height $z$ using 
\begin{equation}
\sqrt{D_z^2+z^2}+\sqrt{\left(2\sqrt{2}\,W/\pi-D_z + O(\bar{u}_c^2)\right)^2+z^2} = W \,.
\end{equation}
The distance $\Delta$, induced by the rotation of the projection of the now-curved material line that began as a generator in the U state (Figure \ref{trig}(b)), can be approximated using similar triangles: 
\begin{equation}
\frac{\sqrt{D_z^2+z^2}}{D_z} = \frac{W/2}{\Delta + \sqrt{2}\,W/\pi + O(\bar{u}_c^2)} \,.
\end{equation}
As this projection is perpendicular to the edges, 
another similar triangle relation yields
\begin{equation}
	\frac{u_c}{2\sqrt{2}\,W/\pi + O(\bar{u}_c^2)} = \frac{\Delta}{L/2}\,.
\end{equation}
All of this allows us to derive an expression for the critical displacement 
\begin{equation}
  \bar{u}_c (\bar{L}) \approx \frac{4\sqrt{2}\bar{\Delta}}{\pi\bar{L}} \approx 0.06/\bar{L} \,,
  \label{eq:uc}
\end{equation}
used in Figure \ref{composite}(b).

\newpage

 \section{Measurement of crumple sizes}\label{crumplemeasure}

Given the uncertainty and limited range of the crumple size data in Figure \ref{composite}(d), either the expected $1/3$ or an unexpected $2/5$ scaling would be reasonable conclusions, and we cannot definitively exclude a slightly stronger scaling of $1/2$.  Linear least squares fits 
excluding the most uncertain/ambiguous end points\footnote{The smallest thickness sheet is semi-transparent and the largest thickness sheet, used only for the diameter measurements, is black, making reliable data difficult or impossible to obtain.} yield slopes for crumple diameter 0.38 (0.4 retaining smallest and largest thicknesses), nucleation forward 0.35 (0.45 retaining smallest thickness), nucleation return 0.38 (0.36 retaining smallest thickness). 

\begin{figure}[h]
\centering
        \includegraphics[width=0.3\linewidth]{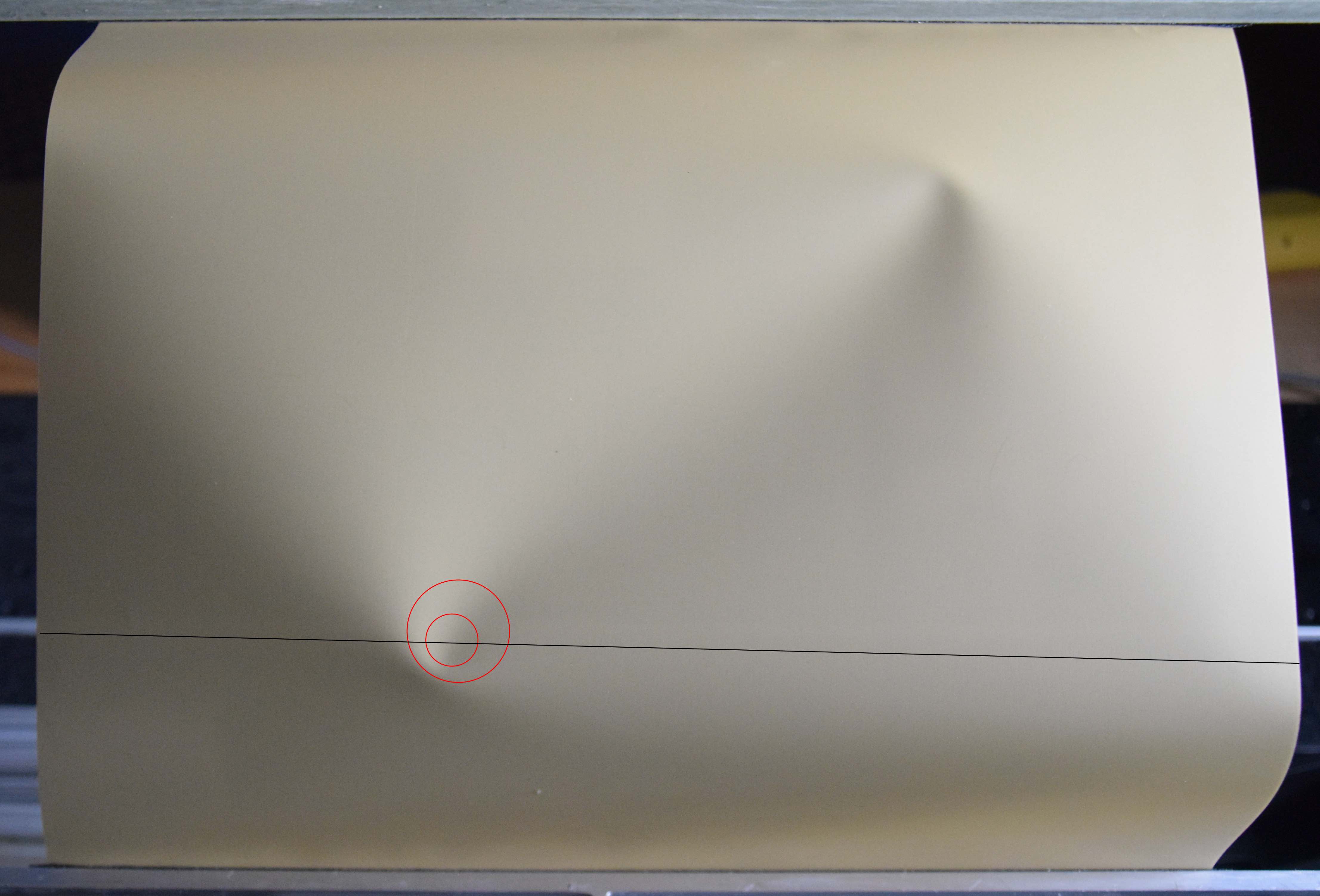}
   \caption{ 
   Example of a crumple measurement (red circles), for a thickness $\bar h = 4\times 10^{-4}$ sheet. 
   The dark line is for calibration. }
    \label{measurement} 
\end{figure}

Crumple diameter data are taken from single crumples induced as part of a well-separated O-valley pair.  
The crescent shape is too diffuse to employ the method of \cite{Liang08}. 
Instead we measure an image that is approximately an orthogonal projection of the crumple, whose crescent curve is essentially planar, by circumscribing and inscribing two circles along the ridge and taking the arithmetic mean of their diameters, with conservative error bars denoting the entire range between the two values. 
An example is shown in Figure \ref{measurement}. 
One set of data was employed, with at least two independent determinations of circle sizes and positions, followed by a group consensus.  The initial circle determinations were performed three times, and the method was found to be reproducible, with only small differences in the results across trials or across the group.

\bibliographystyle{unsrt}

\end{document}